\documentclass{elsart3p}

 \usepackage{epsfig}

\usepackage{amssymb}

\begin{document}

\begin{frontmatter}

\title{Magnetic field induced rotation of the d-vector in Sr$_2$RuO$_4$}
 
\author[a]{J. F. Annett}
\author[b]{G. Litak}
\author[a]{B. L. Gyorffy}
\author[c]{K. I. Wysoki\'nski}
\address[a]{H. H. Wills Physics Laboratory, University of Bristol,  
Tyndall Ave, BS8-1TL,UK}
\address[b]{Department of Mechanics, Technical University of Lublin,
Nadbystrzycka 36, PL-20-618 Lublin, Poland}
\address[c]{Institute of Physics, M. Curie-Sk\l{}odowska University,
Radziszewskiego 10, PL-20-031 Lublin, Poland}

\begin{abstract}
The superconductor Sr$_2$RuO$_4$ is widely believed to be a spin triplet system with
a chiral order parameter analogous to the A phase of superfluid helium-3. The best
evidence for this pairing state is that the Knight shift or spin susceptibility measured
in neutron scattering is constant below T$_c$, unlike in a spin-singlet superconductor.
The original Knight shift and neutron scattering measurements were performed for
magnetic fields aligned in the ruthenate a-b plane. These would be consistent with
a triplet d-vector $d(\vec{k})$ aligned along the c-axis. However recently the Knight shift
for fields along c was also found to be constant below T$_c$, which is not expected
for this symmetry state. In this paper we show that while spin-orbit interaction
stabilises the c-axis oriented d-vector, it is possible that only a very small external
B field may be sufficient to rotate the d-vector into the a-b plane. In this case the
triplet pairing model remains valid. We discuss characteristics of the transition and
the prospects to detect it in thermodynamic quantities.
\end{abstract}

\begin{keyword}
 
ruthenate superconductors \sep pairing symmetry \sep d-vector rotation

\PACS 74.20.-z \sep 74.20.Rp \sep 74.70.Pq \sep 74.25.Bt
\end{keyword}
\end{frontmatter}

\section{Introduction}
\label{intro}
 
Strontium ruthenate is an intriguing low T$_c$ 
superconductor \cite{maeno2001}. 
It is widely believed to have  spin triplet  order 
parameter. The orbital symmetry 
of it, however, is still unknown. Knight shift \cite{ishida1998} and spin 
susceptibility \cite{duffy2000}
being constant below $T_c$ point to the chiral state with d-vector 
$\mathbf{d}(\mathbf{k}) \sim  (k_x + i k_y ) \hat{\mathbf{e}}_z$. 
This state (called (a) in the following) is also consistent with the $\mu$-SR
experiments which show spontaneous time reversal symmetry breaking
at $T_c$ \cite{luke1998}. It was recently confirmed in phase sensitive
experiments \cite{Nelson04}. 

The original Knight shift measurements were performed for
magnetic fields aligned in the ruthenate a-b plane.  Recently 
the Knight shift  \cite{murakawa2004} has  been measured  
in magnetic field  parallel to the c-axis.   In contradiction
to expectations it was found 
that the susceptibility is \emph{also
unchanged} from the normal state value below $T_c$. 
It was noted\cite{murakawa2004} that this result would be consistent with 
the assumption that the ${\bf d}$ vector
rotates away from the ${\bf \hat{e}_z}$ direction under the influence of
the external field.  However a simple
rotated state, such as $\mathbf{d}(\mathbf{k}) \sim  (k_x + i k_y ) \hat{\mathbf{e}}_x$,
is not allowed in tetragonal symmetry. Instead, the symmetry allowed states 
 $ {\bf d}({\bf k}) = (\sin{k_x},\sin{k_y},0)$
 and $ {\bf d}({\bf k}) = (\sin{k_y},-\sin{k_x},0)$
 (which below we refer to as $(b)$ and $(c)$ respectively) would 
also explain the data as the spin susceptibility for either of these states is
of the form \cite{annett1990}
\begin{equation}
  \hat{\chi}_s(T) =  (\chi_n/2)  
\rm{ diag} \left( \begin{array}{ccc} 1+Y(T), 1+Y(T), 2 \end{array}\right), 
\end{equation}
corresponding to a constant spin susceptibility for c-axis fields  
consistent with the results of Murakawa {\it et al.} \cite{murakawa2004}.

It is the purpose of this work to study in some detail the scenario
of d-vector rotation in the c-axis magnetic field. 
Specifically, we shall study the possible effect of the weak 
external (Zeeman only) magnetic field on the symmetry of the  
lowest energy pairing state.
 
To this end we use a recently proposed   \cite{annett2002}
phenomenological  three-band,
three dimensional  model with a realistic band structure \cite{annett2002}. 
It takes into account  
the three ruthenium $t_{2g}$ orbitals 
$xz$, $yz$ and $xy$.  The hopping integrals $t_{mm^{ \prime}}(ij)$ for 
orbitals $m$ and $m^\prime$ and Ru sites $i$ and $j$ as well as the
site energies $ \varepsilon_m$ are fitted to reproduce the experimentally
determined Fermi surface  \cite{mackenzie1996}. The
effective spin and orbital dependent attractive Hubbard  parameters   
$U_{mm^{ \prime}}^{ \alpha \beta, \gamma \delta}(ij)$  are chosen to give
correct transition temperature.
Without spin - orbit coupling  the   model with 
only two values of interaction parameters: 
an in - plane $U_{\parallel}=U$ and   out - of - plane $U_{\perp}$
 results in the chiral state 
 $(a)$, as the ground state of the system. 
 
\section{Results}
\label{res}
For spin-independent interactions 
the alternative pairing states $(b)$ and $(c)$  are the 
ground state for any value of $\lambda>0$. 
To get the state (a) as the ground state   
in the presence of non-zero spin-orbit interaction $\lambda$ 
one has to take into account that the 
spin-orbit coupling leads to a small spin dependence of the effective pairing
interaction \cite{annett2006}.  
Choosing $U'_{\parallel}\equiv U^{\uparrow\downarrow}$
 about  1\% larger than $U \equiv U^{\uparrow\uparrow}=U^{\downarrow\downarrow}$
 is sufficient to stabilize the chiral state even for
large spin orbit coupling.  
\begin{figure}[tbp]
\epsfig{file=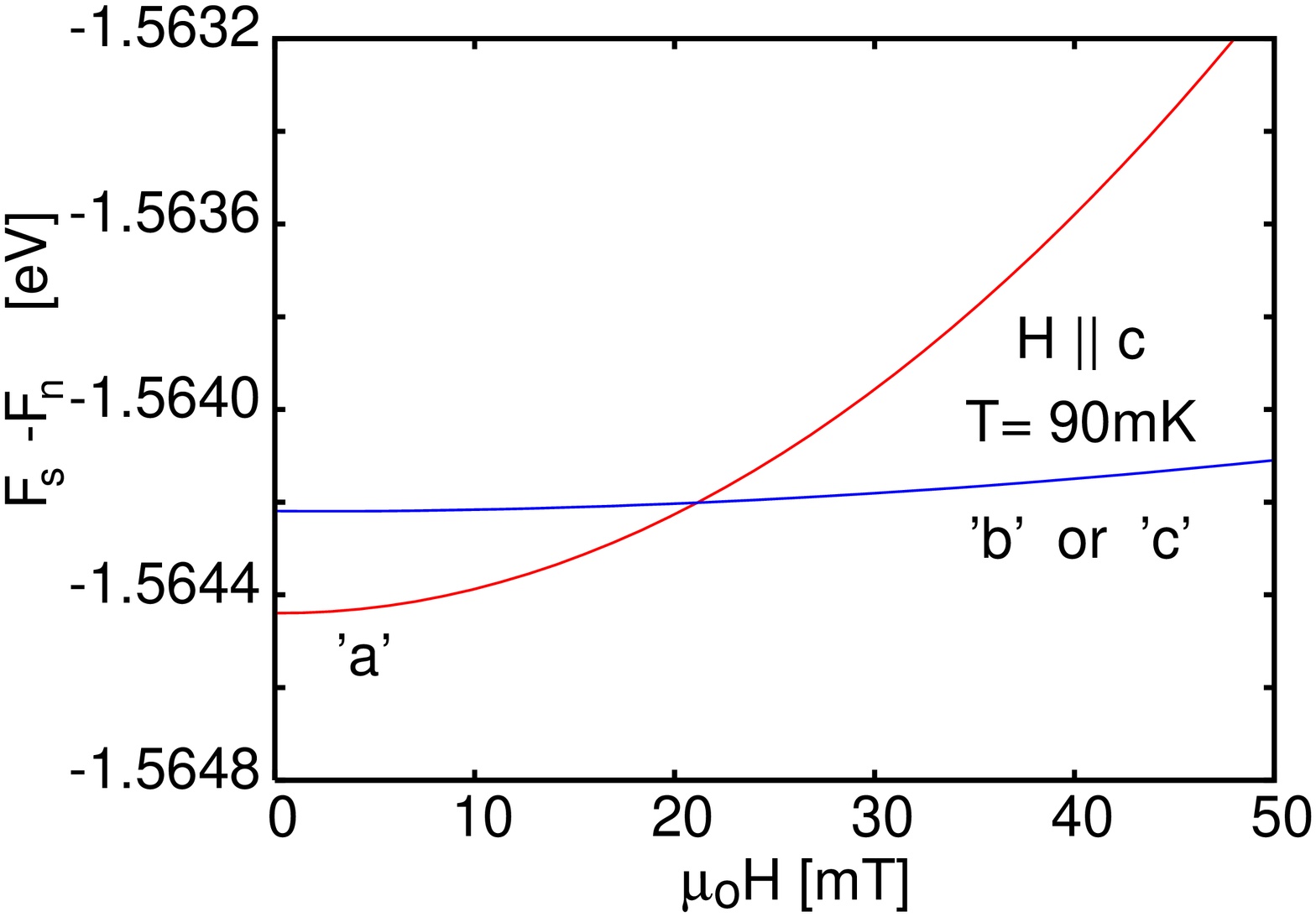,width=6.0cm,angle=-90}

\vspace{-5.4cm}

\hspace{2.2cm}
\epsfig{file=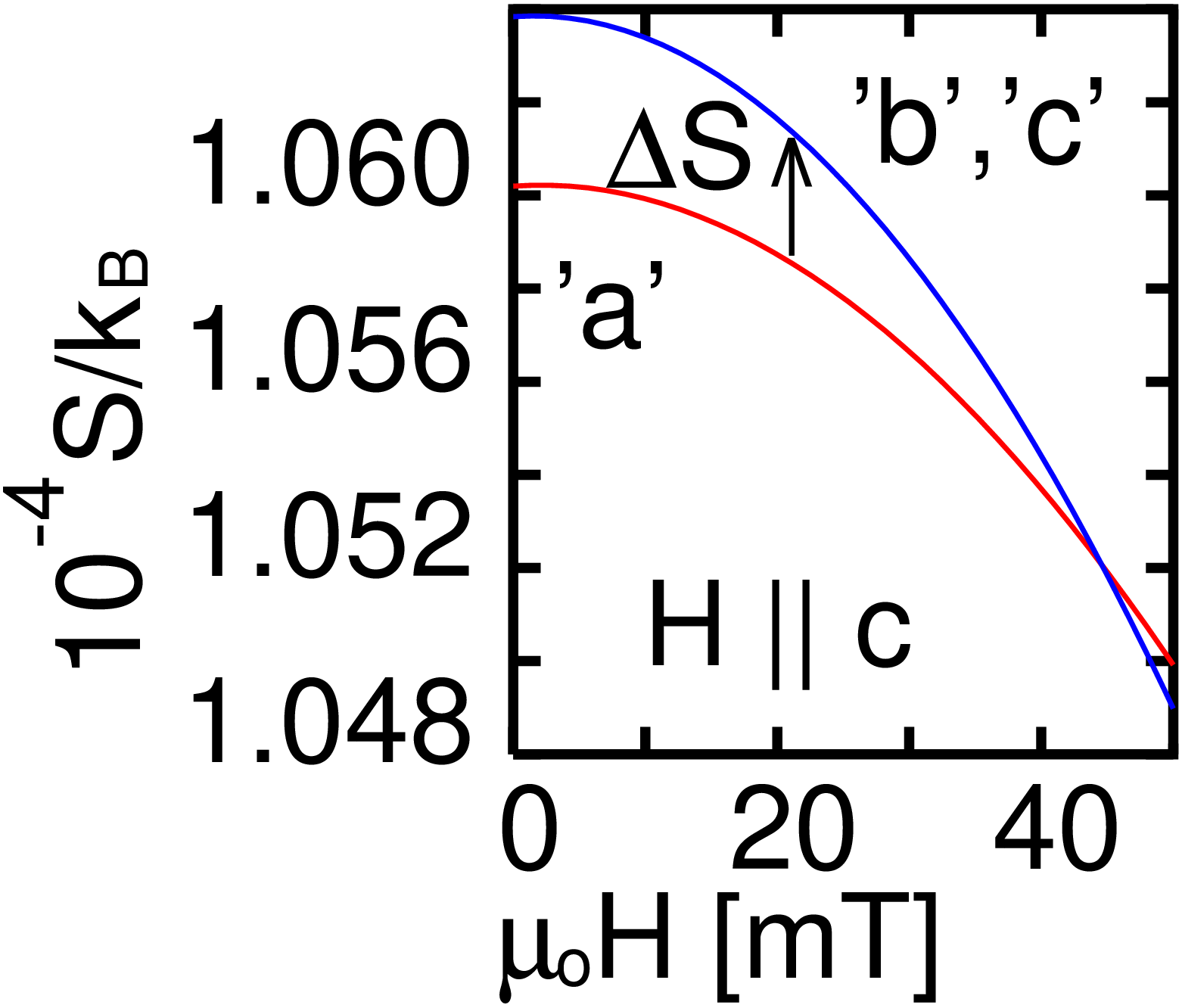,width=2.5cm,angle=-90}

\vspace{3.00cm}
\caption{Condensation free energy at $T=90$mK for three triplet order parameters of
different symmetries, $(a)$, $(b)$ and $(c)$ as a function of
  external field $H
\parallel c$. Here $\lambda=0.02t$ where $t=80$meV is the effective 
$\gamma$ band nearest neighbor hopping integral, and $U'=1.0012U$
in the notation of Ref.~\cite{annett2006}.
\label{fig1}}

\end{figure}
In Fig.~\ref{fig1} we show the free energies of the $(a)$ $(b)$ and $(c)$
symmetry pairing states as a function of external magnetic field
$H\parallel c$.  
The model parameters were chosen to make the chiral state $(a)$ stable
at zero field. One can see that the chiral phase 
increases its free energy as the field is increased, until
at a certain critical field the $(b)$ or $(c)$ solutions
 become more stable. Therefore we
expect a ``spin flop'' type phase transition 
from a d-vector oriented along the $c$-axis
to one where the d-vector lies in the $a-b$ plane. 
 A similar rotation is observed in the bulk superfluid state of $^3He$-A  
 where the d-vector  rotates continuously
to remain perpendicular to the applied field \cite{vollhardt1990}.  

In the case of Sr$_2$RuO$_4$
 the transition is not simply a rotation
of the chiral d-vector but is also a transition from 
a chiral to non-chiral pairing state of different symmetry. 
The two distinct solutions shown in Fig.~\ref{fig1} 
(note that $(b)$ and $(c)$ are essentially degenerate)
have different entropies, as shown in the figure 
inset, and hence this Freedericksz-like \cite{vollhardt1990} spin-flop  is 
a first order thermodynamic phase transition. 
It has not yet been observed,
because the specific heat jump at this transition is 
very small in comparison with the main jump at 1.5K, as our
(unpublished) calculations indicate.

{\bf Acknowledgements:} This work has been partially supported 
by KBN grant No. 2P03B06225 and NATO Collaborative Linkage Grant No. 979446.

\end{document}